\documentclass[aps,prb,reprint,groupedaddress,showpacs]{revtex4-1}

\usepackage{graphicx}
\usepackage{dcolumn}
\usepackage{bm}

\begin{document}


\title{Non-reciprocal light diffraction by a vortex magnetic particle}


\author{E.A.Karashtin}
 \email{eugenk@ipmras.ru}
\affiliation{Institute for Physics of Microstructures RAS, Nizhny Novgorod 603950, GSP-105, Russia}


\date{\today}

\begin{abstract}
We report a theoretical study of light diffraction by a spherical magnetic particle with a vortex magnetization distribution. It is shown that the intensity of the diffracted light involves a non-reciprocal contribution. This contribution depends on the vorticity of particle magnetization. It appears due to the excitation of an electric quadrupole, magnetic dipole and the addition to the electric dipole moment in the particle, that depend on the particle magnetization vorticity. The estimation of the non-reciprocal contrbution for a cobalt particle and two linear polarizations of the incident light fits the data of recent experimental studies in the lattice of triangle magnetic particles by an order of magnitude.
\end{abstract}


\maketitle


\section{Introduction\label{Intro}}
Remarkable achievements in micro- and nano- technologies open new possibilities in fabrication of optical media with novel interesting properties. Photonic three-dimensional crystals exhibit extraordinary optical effects, such as photonic band gap \cite{John87,Yabl87}, negative refractive index \cite{Podol03,Smith04}, and ultraslow light propagation \cite{Scal96}. Surface plasmons in planar metal structures lead to enhanced magneto-optical effects \cite{Belo11,Gitt07,Sap11}, extraordinary light transmission through subwavelength holes \cite{Genet07}, and effective generation of the second harmonic \cite{Anc03,Lamb97}. Interesting objects actively investigated in the past decade are planar chiral structures \cite{Papa03,Vall03,Schw03,Zhang06,Pros05}. Such a structure is ordinarily a two-dimensional regular lattice of non-magnetic particles that possesses the time and spatial inversion symmetries \cite{Bass88}. Nevertheless, light transmission through a planar chiral structure experiences polarization rotation which is different for the light traveling in opposite directions due to the absence of mirror symmetry and strong anisotropy in the orientation of the structure elements. The term ``non-reciprocal'' is often used for this effect, and some speculations appear on the time-reversal symmetry violation \cite{Schw03,Zhang06,Pros05}. Note that the intensity of light diffracted by a planar chiral structure does not change with a reversal of the propagation direction, in accordance with the reciprocity law. The non-reciprocal intensity effects can appear only in the presence of a magnetic field \cite{Groen62,Baran77,Krich98} or in magnetic systems \cite{Brown63,Shel92,Remer84,Yama74,Popk98}. Such effects have been observed recently in a two-dimensional lattice of magnetic vortices \cite{Udal12}. The particles had a triangular shape, which allowed manipulation of their vorticity by an uniform external magnetic field, thus making it identical for all particles.

The problem of scattering of electromagnetic waves by a particle of an arbitrary size with gyrotropic dielectric permittivity and magnetic permeability has been addressed in a general form \cite{Ford78} and for magnetic particles specifically \cite{Lin04, Tar04}. However, the gyrotropy terms are always supposed to be constant over the particle, which corresponds to uniform magnetization. This paper is devoted to theoretical calculations of the diffraction of light by a magnetic vortex that turns out to be a non-uniform magnetic particle. Our theory describes the mechanisms underlying the effect of non-reciprocal light scattering by such a particle and also allows us to estimate its magnitude and compare the theoretical data with the experiment \cite{Udal12}. Note that this effect is assumed to be simply summed over the lattice of particles (i.e., the collective phenomena are neglected), thus making it possible to focus on the light diffraction by a single particle.

In Section~\ref{Phenomenology} we consider some phenomenological arguments in favor of the non-reciprocal light diffraction by a vortex particle. Then a microscopic model of the effect is proposed in the approximation of a spherical particle that is small compared to the wavelength. The main assumptions are outlined in Section~\ref{Assump}. Further, we consider two approaches to the problem. A simple Born approximation (permittivity is close to unity) that helps reveal the existence of the effect is described in Section~\ref{Born}. Yet, within this approximation the effect occurs only for one linear polarization of the incident light. Section~\ref{Perturb} is devoted to calculation for the arbitrary permittivity, based on the first-order perturbation theory with respect to the particle size-to-wavelength ratio. We show that the non-reciprocal term appears due to excitation of the electric quadrupole, magnetic dipole and a small addition to the electric dipole moment, that emit light interfering with the main magnetization-independent electro-dipole radiaton. Finally, we perform simple estimations for the parameters of cobalt and an appropriate particle size and compare our calculations to the experimental results \cite{Udal12}. A summary of our results is given in Section~\ref{Sum}.

\section{Phenomenological Considerations\label{Phenomenology}}

We begin with some phenomenological arguments in favor of the non-reciprocal light diffraction by the vortex particle \cite{Udal12}. If we consider the light scattering cross-section summed over the polarizations of incident and diffracted light, the reciprocity law takes a simple form
\begin{equation} \label{Eq_11}
\sigma \left(\bf{k},\bf{k}',\bf{M}\left(\bf{r}\right)\right)=\sigma \left(-\bf{k}',-\bf{k},-\bf{M}\left(\bf{r}\right)\right),
\end{equation}
here $\sigma $ is the differential cross-section for the diffracted light, $\bf{k}$ and $\bf{k}'$ are the wave vectors of the incident and diffracted  beams, $\bf{M}\left(\bf{r}\right)$ is the magnetization spatial distribution. The term ``non-reciprocal effect'' implies the following inequality:
\begin{equation} \label{Eq_12}
\sigma \left(\bf{k},\bf{k}',\bf{M}\left(\bf{r}\right)\right)\ne \sigma \left(-\bf{k}',-\bf{k},\bf{M}\left(\bf{r}\right)\right),
\end{equation}
which can be transformed using Eq.~(\ref{Eq_11}) to
\begin{equation} \label{Eq_13}
\sigma \left(\bf{k},\bf{k}',\bf{M}\left(\bf{r}\right)\right)\ne \sigma \left(\bf{k},\bf{k}',-\bf{M}\left(\bf{r}\right)\right),
\end{equation}
thus making it possible to observe the effect by inverting the magnetization direction instead of swapping the source and the detector \cite{Udal12}.

For systems without center of inversion the scattering cross-section may contain the term $\left(\left(\bf{k}+\bf{k}'\right)\cdot \bf{C}\right)$, where $\bf C$ is vector. Being linear in the wavevector, this term leads to the non-reciprocal effects described by (\ref{Eq_12},~\ref{Eq_13}). The $\bf{C}$ vector should be a {\it polar vector} that also changes its sign under the time reversal. For a magnetic scatterer of a centrosymmetrical shape and material $\bf{C}$ can be chosen in the simplest form $\bf{C}=\alpha \left\langle \left[\bf{r}\times \bf{M}\left(\bf{r}\right)\right]\right\rangle$ which is a toroidal moment of the particle associated with the magnetic vorticity \cite{Pros08}, the brackets mean the spatial averaging over the scatterer, $\alpha$ is a constant. According to these considerations, the diffraction of unpolarized light by a particle with the vortex magnetization distribution is non-reciprocal and the scattering cross-section has a contribution depending on the vorticity:
\begin{equation} \label{Eq_14}
\sigma \left(\bf{k},\bf{k}',\bf{M}\left(\bf{r}\right)\right)=...+\alpha \left(\left(\bf{k}+\bf{k}'\right)\cdot \left\langle \left[\bf{r}\times \bf{M}\left(\bf{r}\right)\right]\right\rangle \right).
\end{equation}

The weak point of this phenomenological consideration is the polarization dependence of the non-reciprocal effect. Indeed, the existence of the effect for unpolarized light means it exists for at least one linear polarization, but the question about the contribution of different polarizations to it arises. However, the experiments currently are performed exactly for a linearly polarized light \cite{Udal12}.

\section{Main Assumptions And Definitions\label{Assump}}

\begin{figure}[t]
\includegraphics[width=3.25in, keepaspectratio=true]{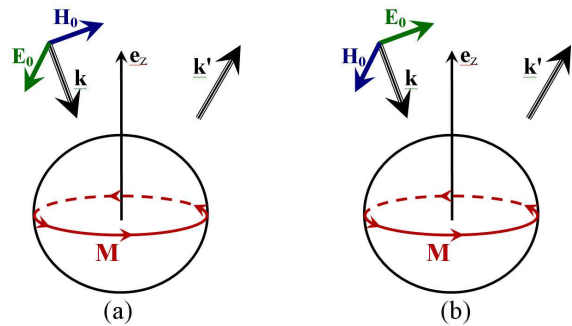}
\caption{\label{Fig_1} (Color online) The geometry of light diffraction by a vortex magnetic particle. (a) The incident light is s-polarized. (b) The incident light is p-polarized.}
\end{figure}

As has already been mentioned, we assume that the diffracting particle has a spherical shape with radius $a$. The incident wave is assumed to be monochromatic (all fields change in time as $e^{i \omega t}$, $\omega$ is the wave circular frequency) and plane. Its wavelength $\lambda$ is much bigger than $a$:
\begin{equation} \label{Eq_20}
\lambda >> a.
\end{equation}
The particle has a vortex magnetic moment (see Fig.~\ref{Fig_1})
\begin{equation} \label{Eq_21}
\bf{M} = \bf{e}_{\phi} = \bf{e}_{y} cos\phi - \bf{e}_{x} sin\phi.
\end{equation}
Here $\bf{e}_{\phi}$ is the unit vector in spherical coordinates, $\bf{e}_{x}$ and $\bf{e}_{y}$ are the unit vectors in Cartesian coordinates, $\phi$ is a spherical coordinate, magnetic moment $\bf{M}$ is normalized to unity.

The magnetic permeability tensor is unit, while the dielectric permittivity tensor is assumed to have a locally gyrotropic form:
\begin{equation} \label{Eq_22}
\left(\hat{\epsilon}\right)_{j l} = \varepsilon \delta_{j l} + i \gamma e_{j l m} M_{m},
\end{equation}
where $\delta_{j l}$ is the Kronecker delta, $e_{j l m}$ is the completely antisymmetric tensor (the Levi-Civita tensor). Since $\bf{M}$ is normalized to unity, the $\gamma$ coefficient includes its magnitude. Expression (\ref{Eq_22}) is valid if the magnitude of quasiclassic electron oscillations is small compared to the scale of the magnetic moment variation. In our case it is equal to the particle radius $a$, hence, we get applicability criterion
\begin{equation} \label{Eq_23}
\frac{e E_{0} \lambda}{m c^{2}} \lambda << a,
\end{equation}
where $e$ is the absolute electron charge, $m$ is the electron mass, $E_{0}$ is the magnitude of the electric field of the incident wave, $c$ stands for the light velocity. The first term in the left-hand part of (\ref{Eq_23}) is the ratio of the energy the electron gains while oscillating in the electric field of the wave, to its rest energy. In the optical range this ratio is very small for the existing sources, which allows simultaneous fulfillment of conditions (\ref{Eq_20}) and (\ref{Eq_23}).

Another assumption used in our symmetry considerations (Section~\ref{Phenomenology}) and elsewhere throughout the paper is that the diffracted wave is plane. In fact, the wave diffracted by a small particle is spherical at a large distance from the latter. But on a scale much smaller than the distance between the particle and the measuring point the wave may be considered plane \cite{Landau8}. Its intensity, indeed, depends on the distance from the particle.

Since the experiments on light diffraction by the lattice of magnetic vortices have been carried out for the linear polarizations of incident light, we also use these polarizations in theoretical calculations. The following designations for the linear polarizations are used throughout the paper: the wave polarized so that its electric field vector lies in the plane defined by the $\bf{k}$ vector and the axis of the magnetic vortex $\bf{e}_{z}$ is termed p-polarized (see Fig.~\ref{Fig_1}); the wave with the electric field perpendicular to that plane is s-polarized.

One more important approximation consists in restricting ourselves to the first order in the magnetic moment, which mathematically corresponds to the first order in a small parameter $\gamma / \varepsilon$ (see Eq.~(\ref{Eq_22})). Our choice of approximation is determined by both the fact that the phenomenon described by Eq.~(\ref{Eq_14}) is linear in $\bf{M}$ and that $\gamma / \varepsilon$ is a very small parameter for the existing ferromagnets \cite{Ber99,Deh01}.

\section{The Born Approximation\label{Born}}

The simplest calculation is based on the assumption that permittivity tensor $\hat{\epsilon}$ is almost equal to unit tensor $\hat{1}$:
\begin{equation} \label{Eq_31}
\hat{\epsilon} = \hat{1} + \hat{\delta \epsilon}.
\end{equation}
The Maxwell equations can be solved in this case for a scattering particle of an arbitrary shape \cite{Landau8}. According to the well-known theory, the electric field of diffracted wave $\bf{E_{scat}}$ can be written as
\begin{equation} \label{Eq_32}
\bf{E_{scat}} = -\frac{e^{i k' R_{0}}}{4 \pi R_{0}} \left[\bf{k'} \times \left[\bf{k'} \times \int_{V} \hat{\delta \epsilon} \bf{E}_{0} e^{i \bf{q} \bf{r}} dV\right]\right].
\end{equation}
Here $\bf{E_{0}}$ is the electric field of the incident wave, $\bf{q}$ stands for $\bf{k} - \bf{k'}$, $V$ is the volume of a scattering particle, $R_{0}$ is the distance between the scattering particle and the point where $\bf{E_{scat}}$ is measured. According to our assumptions (Section~\ref{Assump}), $R_{0}$ is much bigger than the linear dimension of scatterer $V^{1/3}$. 

We can now use (\ref{Eq_22}) and (\ref{Eq_31}) to determine $\hat{\delta \epsilon}$ and calculate the diffracted electric field. It is convenient to use the matrix form to represent the diffraction coefficients for different linear polarizations:
\begin{equation} \label{Eq_33}
\hat{S} = \frac{k_{0}^{2}}{4 \pi R_{0}} \left(
\begin{array} {cc}
S_{ss} & S_{sp} \\ S_{ps} & S_{pp}
\end{array}
\right),
\end{equation}
here the diagonal terms represent the diffraction without a change of polarization, while the off-diagonal terms stand for the diffraction to another polarization. The term $k_{0}$ in Eq.~(\ref{Eq_33}) is the absolute value of the wave vector $k_{0} = \omega / c$. The matrix components in (\ref{Eq_33}) are
\begin{equation} \label{Eq_34}
S_{ss} = \int_{V} \left(\varepsilon - 1 \right)e^{i \bf{q} \bf{r}} dV
\end{equation}
\begin{equation} \label{Eq_35}
S_{sp} = i \gamma \left(\bf{k} \cdot \int_{V} \bf{M\left(\bf{r}\right)} e^{i \bf{q} \bf{r}} dV \right)
\end{equation}
\begin{equation} \label{Eq_36}
S_{ps} = -i \gamma \left(\bf{k'} \cdot \int_{V} \bf{M\left(\bf{r}\right)} e^{i \bf{q} \bf{r}} dV \right)
\end{equation}
\begin{eqnarray} \label{Eq_37}
S_{pp} = \left(\bf{k} \cdot \bf{k'} \right) \int_{V} \left(\varepsilon - 1 \right)e^{i \bf{q} \bf{r}} dV - \nonumber\\
- i \gamma \left(\left[\bf{k} \times \bf{k'}\right] \cdot \int_{V} \bf{M\left(\bf{r}\right)} e^{i \bf{q} \bf{r}} dV \right)
\end{eqnarray}

When analyzing Eqs~(\ref{Eq_34})-(\ref{Eq_37}) we should first of all note that the linear polarizations do not mix without magnetization. Since we are interested in the intensity effect that is linear in $\bf{M}$, only the diagonal terms of matrix (\ref{Eq_33}) should be taken into account. The matrix component $S_{ss}$ does not depend on the magnetic moment, hence, we come to a conclusion that the s-polarized light diffraction does not feature non-reciprocal properties, while for the p-polarized light there is a non-reciprocal contribution in the diffraction intensity.

Finally, taking into account Eq.~(\ref{Eq_20}), we can expand the term $e^{i \bf{q} \bf{r}}$ into Fourier series and neglect all but the first two terms ($e^{i \bf{q} \bf{r}} \approx 1 + i \bf{q} \bf{r}$). These correspond to the electrodipole and electroquadrupole terms that are of the zero and the first order in $a/\lambda$. Calculation of the intensity for $\bf{M}$ as given by (\ref{Eq_21}) leads to
\begin{equation} \label{Eq_38}
I_{s} = \frac{k_{0}^{4} a^{6}}{9 R_{0}^{2}} \left|\varepsilon - 1\right|^{2}
\end{equation}
\begin{eqnarray} \label{Eq_39}
I_{p} = \frac{k_{0}^{4} a^{6}}{9 R_{0}^{2}} \left( \left|\varepsilon - 1\right|^{2} \left(\bf{n} \cdot \bf{n'}\right)^{2} - \left(\left(\bf{n} + \bf{n'}\right) \cdot \bf{e}_{z}\right) \times \right. \nonumber\\
\left. \times \frac{3 k_{0} a}{16} \left(\left(\bf{n} \cdot \bf{n'}\right) - \left(\bf{n} \cdot \bf{n'}\right)^{2}\right) Re\left(\left(\varepsilon - 1\right)^{*} \gamma\right)\right)
\end{eqnarray}
Here $I_{p}$ and $I_{s}$ are the intensities of the diffracted light for p and s polarizations of incident light, $\bf{n}$ and $\bf{n'}$ are the wave vectors of the incident and diffracted waves normalized to unity: $\bf{n} = \bf{k} / k_{0}$, $\bf{n'} = \bf{k'} / k_{0}$, $\bf{e}_{z}$ is the unit vector along the vortex axis.

In order to find the intensity of diffracted light for the unpolarized incident light (we label it $I_{np}$), the intensity should be averaged over the polarizations. Formally, it leads to $I_{np} = \frac{1}{2} \left(I_{s} + I_{p}\right)$, where $I_{np} \sim \sigma$. From Eqs~(\ref{Eq_38},~\ref{Eq_39}) it is clear that the above calculation is consistent with the phenomenological considerations.

In the conclusion of this section, we point out that the non-reciprocal light diffraction was of the same order of magnitude for both s and p polarizations of incident light in the experiment\cite{Udal12}. But within the used approach the non-reciprocal contribution exists only for the p-polarized light. In the next section we propose a more general theory that has no such shortcoming.

\section{Perturbation Theory\label{Perturb}}

A more exact solution of the problem is gained through the perturbation theory based on condition (\ref{Eq_20}). To account for the electroquadrupole and magnetodipole terms, we need to expand the solution up to the first order in $a/\lambda$. The correction to the electrodipole term will be gained naturally. Unlike in some simple models \cite{Kur07} we cannot use the quasistatic Maxwell equations here. Hence, our approach is based on the approximations of the Maxwell equations
\begin{eqnarray} \label{Eq_40}
&rot \bf{E} &= i k_{0} \bf{H}, \\ \label{Eq_41}
&rot \bf{H} &= -i k_{0} \hat{\epsilon} \bf{E}, \\ \label{Eq_42}
&div \left(\hat{\epsilon} \bf{E}\right) &= 0, \\ \label{Eq_43}
&div \bf{H} &= 0,
\end{eqnarray}
that follow from the estimation
\begin{eqnarray} \label{Eq_44_1}
rot \bf{E} \sim \bf{E}/a,\\ \label{Eq_44_2}
rot \bf{H} \sim \bf{H}/a.
\end{eqnarray}
Next, we assume that the permittivity tensor is not large. Taking into account that $\gamma/\varepsilon << 1$, we impose the condition on $\varepsilon$: $\left|\varepsilon\right| << \lambda/a$ ($\lambda/a >> 1$ due to (\ref{Eq_20})). Introducing a typical scale of fields in the particle $L \sim \lambda / \sqrt{\left|\varepsilon\right|}$ (for real $\varepsilon$ $L$ is the wavelength in the medium, for imaginary $\varepsilon$ it is the skin depth), we have
\begin{equation} \label{Eq_44_3}
a << L^{2} / \lambda.
\end{equation}
This condition along with Eq.~(\ref{Eq_20}) allows us to derive the equations for the fields of zero and first order in $a/\lambda$ from Eqs~(\ref{Eq_40}-\ref{Eq_43}).

First of all, let us find the electric field inside the particle. As is well-known, in zero order in $a/\lambda$ and in $\gamma/\varepsilon$ the electric and magnetic fields have the form
\begin{eqnarray} \label{Eq_45}
\bf{E}^{0,0} &=& \frac{3}{\varepsilon + 2} \bf{E}_{0}, \\ \label{Eq_46}
\bf{H}^{0,0} &=& \bf{H}_{0} = \left[\bf{n} \times \bf{E}_{0}\right].
\end{eqnarray}
The first upper index in $\bf{E}^{0,0}$ and $\bf{H}^{0,0}$ is used to denote the order in $a/\lambda$, the second index specifies the order in $\gamma/\varepsilon$.

The equation for $\bf{E}^{1,0}$ follows from Eqs~(\ref{Eq_40}) and (\ref{Eq_44_1})
\begin{equation} \label{Eq_47}
rot \bf{E}^{1,0} = i k_{0} \bf{H}^{0,0}.
\end{equation}
By solving this equation with a zero boundary condition at the infinity we arrive at
\begin{eqnarray} \label{Eq_48}
\bf{E}^{1,0} = \frac{i}{2} \left(\left(\bf{k} \cdot \bf{r}\right) \bf{E}_{0} - \left(\bf{E}_{0} \cdot \bf{r}\right) \bf{k}\right) + \nonumber\\
+\frac{5}{2} \frac{i}{2 \varepsilon + 3} \left(\left(\bf{k} \cdot \bf{r}\right) \bf{E}_{0} + \left(\bf{E}_{0} \cdot \bf{r}\right) \bf{k}\right).
\end{eqnarray}

We now calculate the fields linear in the magnetic moment. The equation for $\bf{H}^{0,1}$ is
\begin{eqnarray} \label{Eq_49}
rot \bf{H}^{0,1} &=& 0, \\ \label{Eq_410}
div \bf{H}^{0,1} &=& 0,
\end{eqnarray}
and, taking into account the zero boundary condition at the infinity, we have $\bf{H}^{0,1} = 0$.

Finally, the equations for $\bf{E}^{0,1}$ and $\bf{E}^{1,1}$ are very similar:
\begin{eqnarray} \label{Eq_411}
&rot \bf{E}^{l,1} = 0, \\ \label{Eq_412}
&div \left( \varepsilon \bf{E}^{l,1} + i \gamma \left[\bf{E}^{l,0} \times \bf{M\left(\bf{r}\right)}\right] \right) = 0,
\end{eqnarray}
where $l = 0, 1$. Using Eq.~(\ref{Eq_411}), the solution can be found as $\bf{E}^{l,1} = - \nabla \psi^{l} \left(\bf{r}\right)$, where $\psi$ is the scalar function that satisfies the equation
\begin{equation} \label{Eq_413}
\Delta \psi^{l} = i \frac{\gamma}{\varepsilon} div \left[\bf{E}^{l,0} \times \bf{M}\right].
\end{equation}
The right-hand part of Eq.~(\ref{Eq_413}) may be treated as a charge density distribution. Solving this equation we have
\begin{equation} \label{Eq_414}
\psi^{l} = -\frac{i}{4 \pi} \frac{\gamma}{\varepsilon} \int_{V} \frac{div \left[\bf{E}^{l,0}\left(\bf{r'}\right) \times \bf{M}\left(\bf{r'}\right)\right]}{\left|\bf{r}-\bf{r'}\right|} dV'.
\end{equation}
Substitution of Eqs~(\ref{Eq_45}), (\ref{Eq_48}), (\ref{Eq_21}) in Eq.~(\ref{Eq_414}) yields a formal solution to the problem. Nevertheless, it should be mentioned that the integral over the particle in the right-hand part of Eq.~(\ref{Eq_414}) cannot be expressed as elementary functions.

The next step of our solution procedure is calculation of the electric current density $\bf{j}$ that is related to the electric field \cite{Landau8}:
\begin{equation} \label{Eq_415}
\bf{j} = -\frac{i \omega}{4 \pi} \left(\hat{\epsilon} - \hat{1}\right) \bf{E}.
\end{equation}
Substitution of (\ref{Eq_22}) in (\ref{Eq_415}) yields
\begin{eqnarray} \label{Eq_416}
\bf{j}^{l,0} &=& -\frac{i \omega}{4 \pi} \left(\varepsilon - 1\right) \bf{E}^{l,0}, \\ \label{Eq_417}
\bf{j}^{l,1} &=& -\frac{i \omega}{4 \pi} \left(\varepsilon - 1\right) \bf{E}^{l,1} + \frac{\gamma \omega}{4 \pi} \left[\bf{E}^{l,0} \times \bf{M}\right].
\end{eqnarray}
Knowing the electric current density distribution over the particle, we can calculate the electric field of the diffracted wave far from the diffracting particle \cite{Landau2}:
\begin{equation} \label{Eq_418}
\bf{E}_{scat} = \frac{i k_{0}^{2} e^{i k_{0} R_{0} - i \omega t}}{\omega R_{0}} \int_{V} \bf{j}\left(\bf{r}\right) e^{-i \bf{k'} \bf{r}} dV.
\end{equation}
We expand the expression $e^{-i \bf{k'} \bf{r}}$ in this formula in the Fourier series and keep the first two terms ($e^{-i \bf{k'} \bf{r}} \approx 1 - i \bf{k'} \bf{r}$). This expansion corresponds to the inclusion of the electroquadrupole, magnetodipole and the first-order correction to the electrodipole term. All of these terms have the same order of magnitude in $a/\lambda$. The waves radiated by them all interfere with the zero-order electrodipole radiation. There is no point in separating them.

Finally, for the electric field of the diffracted wave we have
\begin{eqnarray} \label{Eq_418}
\bf{E}_{scat} = \frac{i k_{0}^{2} e^{i k_{0} R_{0} - i \omega t}}{\omega R_{0}} \int_{V} \left(\left(\bf{j}^{0,0} + \bf{j}^{1,0} + \right.\right. \nonumber\\
\left.\left. + \bf{j}^{0,1} + \bf{j}^{1,1}\right) - i \left(\bf{r} \cdot \bf{k'}\right) \left(\bf{j}^{0,0} + \bf{j}^{0,1}\right)\right) dV,
\end{eqnarray}
where $j^{0,0}, j^{1,0}, j^{0,1}, j^{1,1}$ are determined by Eqs~(\ref{Eq_416}, \ref{Eq_417}). We have earlier noted that the integral defining the electric field inside the particle cannot be expressed as elementary functions; however, the double integral that appears in Eq.~(\ref{Eq_418}) is taken by changing the integration order. Thus, $\bf{E}_{scat}$ turns out to be expressed as elementary functions, indeed.

The last step of the calculation is to find the intensity of the diffracted light. Again, for two linear polarizations of the incident light we get
\begin{eqnarray} \label{Eq_419}
I_{s} = \frac{k_{0}^{4} a^{6}}{R_{0}^{2}} \left|\frac{\varepsilon - 1}{\varepsilon + 2}\right|^{2} \left(1 - \left(\left(\bf{n} + \bf{n'}\right) \cdot \bf{e}_{z}\right) \times \right. \nonumber\\
\left. \times \frac{\pi k_{0} a}{32} Re\left(\frac{6 \gamma}{2 \varepsilon + 3} \right)\right)
\end{eqnarray}
\begin{eqnarray} \label{Eq_420}
I_{p} = \frac{k_{0}^{4} a^{6}}{R_{0}^{2}} \left|\frac{\varepsilon - 1}{\varepsilon + 2}\right|^{2} \left(\left(\bf{n} \cdot \bf{n'}\right)^{2} - \left(\left(\bf{n} + \bf{n'}\right) \cdot \bf{e}_{z}\right) \times \right. \nonumber\\
\left. \times \frac{\pi k_{0} a}{32} Re\left(\frac{6 \gamma}{2 \varepsilon + 3} \left(\left(\bf{n} \cdot \bf{n'}\right) + \frac{5}{\varepsilon - 1} \times \right.\right.\right. \nonumber\\
\left.\left.\left. \times \left(\left(\bf{n} \cdot \bf{n'}\right) - \left(\bf{n} \cdot \bf{n'}\right)^{2}\right)\right)\right)\right).
\end{eqnarray}
Eqs~(\ref{Eq_419},~\ref{Eq_420}) manifest that the non-reciprocal contribution exists for both s- and p-polarized incident light. For the p-polarized light they show a somewhat different dependence on the angle between $\bf{n}$ and $\bf{n'}$ from that yielded by the Born approximation. These formulas are completely consistent with the phenomenological considerations (Section~\ref{Phenomenology}).

The perturbation theory that accurately takes into account all of the terms linear in $a/\lambda$ qualitatively agrees with the experiment \cite{Udal12}. To compare the results of our theoretical calculation with the experimental data we performed simple estimations for the parameters of cobalt \cite{Ber99}. The wavelength is $\lambda = 632.8 nm$. At this wavelength cobalt absorbs light quite well, so the permittivity $\epsilon$ is defined by the complex constants $\varepsilon = -12.6 + 22.88i, \gamma = 0.749 - 0.602i$. In the experiment \cite{Udal12} the particles are flat (their thickness is much less than their lateral size). The scale of variation of the internal current density that is important for our calculation is a particle thickness. So we take the particle radius $a = 30 nm$. The incident light propagates along the vortex axis ($\bf{n} = \bf{e}_{z}$) and the diffracted light is deflected from it by $30^{\circ}$. For these parameters we have $\Delta I_{s} / I_{s} \approx 1.12 \cdot 10^{-2}, \Delta I_{p} / I_{p} \approx 1.29 \cdot 10^{-2}$. The experimental value of the effect is $2 \cdot 10^{-3}$. So, we come to a conclusion that although this estimation is very rough it turns out to be a satisfactory fit to the experiment. The difference is apparently attributed to the different shape of the particle and to the violation of the assumption that the permittivity value is not large (see (\ref{Eq_44_3})).

\section{Conclusion\label{Sum}}

We have solved the problem of light diffraction by a spheric magnetic particle with the vortex magnetization distribution up to the first order in the $a/\lambda$ parameter. It should be noted that the approach can easily be expanded to the next-order calculation, but generally it is not necessary in order to gain an insight into the nature of these phenomena. Our results show that the vorticity-dependent non-reciprocal contribution in the intensity of the diffracted light takes place for both s and p polarizations of the incident light. The non-reciprocal term arises due to the interference of the zero-order electrodipole radiation and the first-order electrodipole, electroquadrupole and magnetic dipole radiation that linearly depends on the magnetic moment of the particle. The estimations of the effect value for the parameters of cobalt yield a relative value $\Delta I / I \sim 10^{-2}$ that fits the experiment \cite{Udal12} by an order of magnitude.

\begin{acknowledgments}
The author is thankful to A. A. Fraerman for fruitful discussions. This work was supported by the Russian Foundation for Basic Research (contracts No. 12-02-31393, 12-02-33039), the ``Dynasty'' foundation, and RF Agency for Education of Russian Federation (Rosobrazovanie).
\end{acknowledgments}

\bibliography{scattering}

\providecommand{\noopsort}[1]{}\providecommand{\singleletter}[1]{#1}%
\begin{thebibliography}{35}%
\makeatletter
\providecommand \@ifxundefined [1]{%
 \@ifx{#1\undefined}
}%
\providecommand \@ifnum [1]{%
 \ifnum #1\expandafter \@firstoftwo
 \else \expandafter \@secondoftwo
 \fi
}%
\providecommand \@ifx [1]{%
 \ifx #1\expandafter \@firstoftwo
 \else \expandafter \@secondoftwo
 \fi
}%
\providecommand \natexlab [1]{#1}%
\providecommand \enquote  [1]{``#1''}%
\providecommand \bibnamefont  [1]{#1}%
\providecommand \bibfnamefont [1]{#1}%
\providecommand \citenamefont [1]{#1}%
\providecommand \href@noop [0]{\@secondoftwo}%
\providecommand \href [0]{\begingroup \@sanitize@url \@href}%
\providecommand \@href[1]{\@@startlink{#1}\@@href}%
\providecommand \@@href[1]{\endgroup#1\@@endlink}%
\providecommand \@sanitize@url [0]{\catcode `\\12\catcode `\$12\catcode
  `\&12\catcode `\#12\catcode `\^12\catcode `\_12\catcode `\%12\relax}%
\providecommand \@@startlink[1]{}%
\providecommand \@@endlink[0]{}%
\providecommand \url  [0]{\begingroup\@sanitize@url \@url }%
\providecommand \@url [1]{\endgroup\@href {#1}{\urlprefix }}%
\providecommand \urlprefix  [0]{URL }%
\providecommand \Eprint [0]{\href }%
\providecommand \doibase [0]{http://dx.doi.org/}%
\providecommand \selectlanguage [0]{\@gobble}%
\providecommand \bibinfo  [0]{\@secondoftwo}%
\providecommand \bibfield  [0]{\@secondoftwo}%
\providecommand \translation [1]{[#1]}%
\providecommand \BibitemOpen [0]{}%
\providecommand \bibitemStop [0]{}%
\providecommand \bibitemNoStop [0]{.\EOS\space}%
\providecommand \EOS [0]{\spacefactor3000\relax}%
\providecommand \BibitemShut  [1]{\csname bibitem#1\endcsname}%
\let\auto@bib@innerbib\@empty
\bibitem [{\citenamefont {John}(1987)}]{John87}%
  \BibitemOpen
  \bibfield  {author} {\bibinfo {author} {\bibfnamefont {S.}~\bibnamefont
  {John}},\ }\href@noop {} {\bibfield  {journal} {\bibinfo  {journal} {Phys.\
  Rev.\ Lett.}\ }\textbf {\bibinfo {volume} {58}},\ \bibinfo {pages} {2486}
  (\bibinfo {year} {1987})}\BibitemShut {NoStop}%
\bibitem [{\citenamefont {Yablonovitch}(1987)}]{Yabl87}%
  \BibitemOpen
  \bibfield  {author} {\bibinfo {author} {\bibfnamefont {E.}~\bibnamefont
  {Yablonovitch}},\ }\href@noop {} {\bibfield  {journal} {\bibinfo  {journal}
  {Phys.\ Rev.\ Lett.}\ }\textbf {\bibinfo {volume} {58}},\ \bibinfo {pages}
  {2059} (\bibinfo {year} {1987})}\BibitemShut {NoStop}%
\bibitem [{\citenamefont {Podolskiy}\ \emph {et~al.}(2003)\citenamefont
  {Podolskiy}, \citenamefont {Sarychev},\ and\ \citenamefont
  {Shalaev}}]{Podol03}%
  \BibitemOpen
  \bibfield  {author} {\bibinfo {author} {\bibfnamefont {V.~A.}\ \bibnamefont
  {Podolskiy}}, \bibinfo {author} {\bibfnamefont {A.~K.}\ \bibnamefont
  {Sarychev}}, \ and\ \bibinfo {author} {\bibfnamefont {V.~M.}\ \bibnamefont
  {Shalaev}},\ }\href@noop {} {\bibfield  {journal} {\bibinfo  {journal} {Opt.
  Express}\ }\textbf {\bibinfo {volume} {11}},\ \bibinfo {pages} {735}
  (\bibinfo {year} {2003})}\BibitemShut {NoStop}%
\bibitem [{\citenamefont {Smith}\ \emph {et~al.}(2004)\citenamefont {Smith},
  \citenamefont {Pendry},\ and\ \citenamefont {Wiltshire}}]{Smith04}%
  \BibitemOpen
  \bibfield  {author} {\bibinfo {author} {\bibfnamefont {D.~R.}\ \bibnamefont
  {Smith}}, \bibinfo {author} {\bibfnamefont {J.~B.}\ \bibnamefont {Pendry}}, \
  and\ \bibinfo {author} {\bibfnamefont {M.~C.~K.}\ \bibnamefont {Wiltshire}},\
  }\href@noop {} {\bibfield  {journal} {\bibinfo  {journal} {Science}\ }\textbf
  {\bibinfo {volume} {305}},\ \bibinfo {pages} {788} (\bibinfo {year}
  {2004})}\BibitemShut {NoStop}%
\bibitem [{\citenamefont {Scalora}\ \emph {et~al.}(1996)\citenamefont
  {Scalora}, \citenamefont {Flynn}, \citenamefont {Reinhardt},\ and\
  \citenamefont {Fork}}]{Scal96}%
  \BibitemOpen
  \bibfield  {author} {\bibinfo {author} {\bibfnamefont {M.}~\bibnamefont
  {Scalora}}, \bibinfo {author} {\bibfnamefont {R.~J.}\ \bibnamefont {Flynn}},
  \bibinfo {author} {\bibfnamefont {S.~B.}\ \bibnamefont {Reinhardt}}, \ and\
  \bibinfo {author} {\bibfnamefont {R.~L.}\ \bibnamefont {Fork}},\ }\href@noop
  {} {\bibfield  {journal} {\bibinfo  {journal} {Phys.\ Rev.\ E}\ }\textbf
  {\bibinfo {volume} {54}},\ \bibinfo {pages} {R1078} (\bibinfo {year}
  {1996})}\BibitemShut {NoStop}%
\bibitem [{\citenamefont {Belotelov}\ \emph {et~al.}(2011)\citenamefont
  {Belotelov}, \citenamefont {Akimov}, \citenamefont {Pohl} \emph
  {et~al.}}]{Belo11}%
  \BibitemOpen
  \bibfield  {author} {\bibinfo {author} {\bibfnamefont {V.~I.}\ \bibnamefont
  {Belotelov}}, \bibinfo {author} {\bibfnamefont {I.~A.}\ \bibnamefont
  {Akimov}}, \bibinfo {author} {\bibfnamefont {M.}~\bibnamefont {Pohl}},  \emph
  {et~al.},\ }\href@noop {} {\bibfield  {journal} {\bibinfo  {journal} {Nature
  Nanotechnology}\ }\textbf {\bibinfo {volume} {6}},\ \bibinfo {pages} {370}
  (\bibinfo {year} {2011})}\BibitemShut {NoStop}%
\bibitem [{\citenamefont {Gittis}\ \emph {et~al.}(2007)\citenamefont {Gittis},
  \citenamefont {Papaioannou}, \citenamefont {Patoka} \emph {et~al.}}]{Gitt07}%
  \BibitemOpen
  \bibfield  {author} {\bibinfo {author} {\bibfnamefont {G.}~\bibnamefont
  {Gittis}}, \bibinfo {author} {\bibfnamefont {E.}~\bibnamefont {Papaioannou}},
  \bibinfo {author} {\bibfnamefont {P.}~\bibnamefont {Patoka}},  \emph
  {et~al.},\ }\href@noop {} {\bibfield  {journal} {\bibinfo  {journal} {Phys.\
  Rev.\ Lett.}\ }\textbf {\bibinfo {volume} {98}},\ \bibinfo {pages} {077401}
  (\bibinfo {year} {2007})}\BibitemShut {NoStop}%
\bibitem [{\citenamefont {M.V.Sapozhnikov}\ \emph {et~al.}(2011)\citenamefont
  {M.V.Sapozhnikov}, \citenamefont {S.A.Gusev}, \citenamefont {B.B.Troitskii},\
  and\ \citenamefont {L.V.Khokhlova}}]{Sap11}%
  \BibitemOpen
  \bibfield  {author} {\bibinfo {author} {\bibnamefont {M.V.Sapozhnikov}},
  \bibinfo {author} {\bibnamefont {S.A.Gusev}}, \bibinfo {author} {\bibnamefont
  {B.B.Troitskii}}, \ and\ \bibinfo {author} {\bibnamefont {L.V.Khokhlova}},\
  }\href@noop {} {\bibfield  {journal} {\bibinfo  {journal} {Opt. Lett.}\
  }\textbf {\bibinfo {volume} {36}},\ \bibinfo {pages} {4197} (\bibinfo {year}
  {2011})}\BibitemShut {NoStop}%
\bibitem [{\citenamefont {Genet}\ and\ \citenamefont
  {Ebbesen}(2007)}]{Genet07}%
  \BibitemOpen
  \bibfield  {author} {\bibinfo {author} {\bibfnamefont {C.}~\bibnamefont
  {Genet}}\ and\ \bibinfo {author} {\bibfnamefont {T.}~\bibnamefont
  {Ebbesen}},\ }\href@noop {} {\bibfield  {journal} {\bibinfo  {journal}
  {Nature}\ }\textbf {\bibinfo {volume} {445}},\ \bibinfo {pages} {39}
  (\bibinfo {year} {2007})}\BibitemShut {NoStop}%
\bibitem [{\citenamefont {Anceau}\ \emph {et~al.}(2003)\citenamefont {Anceau},
  \citenamefont {Brasselet}, \citenamefont {Zyss},\ and\ \citenamefont
  {Gadenne}}]{Anc03}%
  \BibitemOpen
  \bibfield  {author} {\bibinfo {author} {\bibfnamefont {C.}~\bibnamefont
  {Anceau}}, \bibinfo {author} {\bibfnamefont {S.}~\bibnamefont {Brasselet}},
  \bibinfo {author} {\bibfnamefont {J.}~\bibnamefont {Zyss}}, \ and\ \bibinfo
  {author} {\bibfnamefont {P.}~\bibnamefont {Gadenne}},\ }\href@noop {}
  {\bibfield  {journal} {\bibinfo  {journal} {Opt. Lett.}\ }\textbf {\bibinfo
  {volume} {28}},\ \bibinfo {pages} {713} (\bibinfo {year} {2003})}\BibitemShut
  {NoStop}%
\bibitem [{\citenamefont {Lambrecht}\ \emph {et~al.}(1997)\citenamefont
  {Lambrecht}, \citenamefont {Leitner},\ and\ \citenamefont
  {Aussenegg}}]{Lamb97}%
  \BibitemOpen
  \bibfield  {author} {\bibinfo {author} {\bibfnamefont {B.}~\bibnamefont
  {Lambrecht}}, \bibinfo {author} {\bibfnamefont {A.}~\bibnamefont {Leitner}},
  \ and\ \bibinfo {author} {\bibfnamefont {F.~R.}\ \bibnamefont {Aussenegg}},\
  }\href@noop {} {\bibfield  {journal} {\bibinfo  {journal} {Appl. Phys. B}\
  }\textbf {\bibinfo {volume} {64}},\ \bibinfo {pages} {269} (\bibinfo {year}
  {1997})}\BibitemShut {NoStop}%
\bibitem [{\citenamefont {Papakostas}\ \emph {et~al.}(2003)\citenamefont
  {Papakostas}, \citenamefont {Potts}, \citenamefont {Bagnall}, \citenamefont
  {Prosvirnin}, \citenamefont {Coles},\ and\ \citenamefont
  {Zheludev}}]{Papa03}%
  \BibitemOpen
  \bibfield  {author} {\bibinfo {author} {\bibfnamefont {A.}~\bibnamefont
  {Papakostas}}, \bibinfo {author} {\bibfnamefont {A.}~\bibnamefont {Potts}},
  \bibinfo {author} {\bibfnamefont {D.~M.}\ \bibnamefont {Bagnall}}, \bibinfo
  {author} {\bibfnamefont {S.~L.}\ \bibnamefont {Prosvirnin}}, \bibinfo
  {author} {\bibfnamefont {H.~J.}\ \bibnamefont {Coles}}, \ and\ \bibinfo
  {author} {\bibfnamefont {N.~I.}\ \bibnamefont {Zheludev}},\ }\href@noop {}
  {\bibfield  {journal} {\bibinfo  {journal} {Phys.\ Rev.\ Lett.}\ }\textbf
  {\bibinfo {volume} {90}},\ \bibinfo {pages} {107404} (\bibinfo {year}
  {2003})}\BibitemShut {NoStop}%
\bibitem [{\citenamefont {Vallius}\ \emph {et~al.}(2003)\citenamefont
  {Vallius}, \citenamefont {Jefimovs}, \citenamefont {Turunen}, \citenamefont
  {Vahimaa},\ and\ \citenamefont {Svirko}}]{Vall03}%
  \BibitemOpen
  \bibfield  {author} {\bibinfo {author} {\bibfnamefont {T.}~\bibnamefont
  {Vallius}}, \bibinfo {author} {\bibfnamefont {K.}~\bibnamefont {Jefimovs}},
  \bibinfo {author} {\bibfnamefont {J.}~\bibnamefont {Turunen}}, \bibinfo
  {author} {\bibfnamefont {P.}~\bibnamefont {Vahimaa}}, \ and\ \bibinfo
  {author} {\bibfnamefont {Y.}~\bibnamefont {Svirko}},\ }\href@noop {}
  {\bibfield  {journal} {\bibinfo  {journal} {Appl. Phys. Lett.}\ }\textbf
  {\bibinfo {volume} {83}},\ \bibinfo {pages} {234} (\bibinfo {year}
  {2003})}\BibitemShut {NoStop}%
\bibitem [{\citenamefont {Schwanecke}\ \emph {et~al.}(2003)\citenamefont
  {Schwanecke}, \citenamefont {Krasavin}, \citenamefont {Bagnall} \emph
  {et~al.}}]{Schw03}%
  \BibitemOpen
  \bibfield  {author} {\bibinfo {author} {\bibfnamefont {A.~S.}\ \bibnamefont
  {Schwanecke}}, \bibinfo {author} {\bibfnamefont {A.}~\bibnamefont
  {Krasavin}}, \bibinfo {author} {\bibfnamefont {D.~M.}\ \bibnamefont
  {Bagnall}},  \emph {et~al.},\ }\href@noop {} {\bibfield  {journal} {\bibinfo
  {journal} {Phys.\ Rev.\ Lett.}\ }\textbf {\bibinfo {volume} {91}},\ \bibinfo
  {pages} {247404} (\bibinfo {year} {2003})}\BibitemShut {NoStop}%
\bibitem [{\citenamefont {Zhang}\ \emph {et~al.}(2006)\citenamefont {Zhang},
  \citenamefont {Potts},\ and\ \citenamefont {Bagnall}}]{Zhang06}%
  \BibitemOpen
  \bibfield  {author} {\bibinfo {author} {\bibfnamefont {W.}~\bibnamefont
  {Zhang}}, \bibinfo {author} {\bibfnamefont {A.}~\bibnamefont {Potts}}, \ and\
  \bibinfo {author} {\bibfnamefont {D.~M.}\ \bibnamefont {Bagnall}},\
  }\href@noop {} {\bibfield  {journal} {\bibinfo  {journal} {J. Opt. A, Pure
  Appl. Opt.}\ }\textbf {\bibinfo {volume} {8}},\ \bibinfo {pages} {878}
  (\bibinfo {year} {2006})}\BibitemShut {NoStop}%
\bibitem [{\citenamefont {Prosvirnin}\ and\ \citenamefont
  {Zheludev}(2005)}]{Pros05}%
  \BibitemOpen
  \bibfield  {author} {\bibinfo {author} {\bibfnamefont {S.~L.}\ \bibnamefont
  {Prosvirnin}}\ and\ \bibinfo {author} {\bibfnamefont {N.~I.}\ \bibnamefont
  {Zheludev}},\ }\href@noop {} {\bibfield  {journal} {\bibinfo  {journal}
  {Phys.\ Rev.\ E}\ }\textbf {\bibinfo {volume} {71}},\ \bibinfo {pages}
  {037603} (\bibinfo {year} {2005})}\BibitemShut {NoStop}%
\bibitem [{\citenamefont {Bassiri}\ \emph {et~al.}(1988)\citenamefont
  {Bassiri}, \citenamefont {Papas},\ and\ \citenamefont {Engheta}}]{Bass88}%
  \BibitemOpen
  \bibfield  {author} {\bibinfo {author} {\bibfnamefont {S.}~\bibnamefont
  {Bassiri}}, \bibinfo {author} {\bibfnamefont {C.~H.}\ \bibnamefont {Papas}},
  \ and\ \bibinfo {author} {\bibfnamefont {N.}~\bibnamefont {Engheta}},\
  }\href@noop {} {\bibfield  {journal} {\bibinfo  {journal} {J. Opt. Soc. Am.
  A}\ }\textbf {\bibinfo {volume} {5}},\ \bibinfo {pages} {1450} (\bibinfo
  {year} {1988})}\BibitemShut {NoStop}%
\bibitem [{\citenamefont {Groenewege}(1962)}]{Groen62}%
  \BibitemOpen
  \bibfield  {author} {\bibinfo {author} {\bibfnamefont {M.~P.}\ \bibnamefont
  {Groenewege}},\ }\href@noop {} {\bibfield  {journal} {\bibinfo  {journal}
  {Molec. Phys.}\ }\textbf {\bibinfo {volume} {5}},\ \bibinfo {pages} {541}
  (\bibinfo {year} {1962})}\BibitemShut {NoStop}%
\bibitem [{\citenamefont {Baranova}\ \emph {et~al.}(1977)\citenamefont
  {Baranova}, \citenamefont {Bogdanov},\ and\ \citenamefont
  {Zeldovich}}]{Baran77}%
  \BibitemOpen
  \bibfield  {author} {\bibinfo {author} {\bibfnamefont {N.~B.}\ \bibnamefont
  {Baranova}}, \bibinfo {author} {\bibfnamefont {Y.~V.}\ \bibnamefont
  {Bogdanov}}, \ and\ \bibinfo {author} {\bibfnamefont {B.~Y.}\ \bibnamefont
  {Zeldovich}},\ }\href@noop {} {\bibfield  {journal} {\bibinfo  {journal}
  {Opt. Commun.}\ }\textbf {\bibinfo {volume} {22}},\ \bibinfo {pages} {243}
  (\bibinfo {year} {1977})}\BibitemShut {NoStop}%
\bibitem [{\citenamefont {Krichevtsov}\ \emph {et~al.}(1998)\citenamefont
  {Krichevtsov}, \citenamefont {Pisarev}, \citenamefont {Rzhevsky} \emph
  {et~al.}}]{Krich98}%
  \BibitemOpen
  \bibfield  {author} {\bibinfo {author} {\bibfnamefont {B.~B.}\ \bibnamefont
  {Krichevtsov}}, \bibinfo {author} {\bibfnamefont {R.~V.}\ \bibnamefont
  {Pisarev}}, \bibinfo {author} {\bibfnamefont {A.~A.}\ \bibnamefont
  {Rzhevsky}},  \emph {et~al.},\ }\href@noop {} {\bibfield  {journal} {\bibinfo
   {journal} {Phys.\ Rev.\ B}\ }\textbf {\bibinfo {volume} {57}},\ \bibinfo
  {pages} {14611} (\bibinfo {year} {1998})}\BibitemShut {NoStop}%
\bibitem [{\citenamefont {Brown}\ \emph {et~al.}(1963)\citenamefont {Brown},
  \citenamefont {Shtrikman},\ and\ \citenamefont {Treves}}]{Brown63}%
  \BibitemOpen
  \bibfield  {author} {\bibinfo {author} {\bibfnamefont {W.~F.}\ \bibnamefont
  {Brown}}, \bibinfo {author} {\bibfnamefont {S.}~\bibnamefont {Shtrikman}}, \
  and\ \bibinfo {author} {\bibfnamefont {D.}~\bibnamefont {Treves}},\
  }\href@noop {} {\bibfield  {journal} {\bibinfo  {journal} {J. Appl. Phys.}\
  }\textbf {\bibinfo {volume} {34}},\ \bibinfo {pages} {1233} (\bibinfo {year}
  {1963})}\BibitemShut {NoStop}%
\bibitem [{\citenamefont {Shelankov}\ and\ \citenamefont
  {Pikus}(1992)}]{Shel92}%
  \BibitemOpen
  \bibfield  {author} {\bibinfo {author} {\bibfnamefont {A.~L.}\ \bibnamefont
  {Shelankov}}\ and\ \bibinfo {author} {\bibfnamefont {G.~E.}\ \bibnamefont
  {Pikus}},\ }\href@noop {} {\bibfield  {journal} {\bibinfo  {journal} {Phys.\
  Rev.\ B}\ }\textbf {\bibinfo {volume} {46}},\ \bibinfo {pages} {3326}
  (\bibinfo {year} {1992})}\BibitemShut {NoStop}%
\bibitem [{\citenamefont {Remer}\ \emph {et~al.}(1984)\citenamefont {Remer},
  \citenamefont {Mohler}, \citenamefont {Grill} \emph {et~al.}}]{Remer84}%
  \BibitemOpen
  \bibfield  {author} {\bibinfo {author} {\bibfnamefont {L.}~\bibnamefont
  {Remer}}, \bibinfo {author} {\bibfnamefont {E.}~\bibnamefont {Mohler}},
  \bibinfo {author} {\bibfnamefont {W.}~\bibnamefont {Grill}},  \emph
  {et~al.},\ }\href@noop {} {\bibfield  {journal} {\bibinfo  {journal} {Phys.\
  Rev.\ B}\ }\textbf {\bibinfo {volume} {30}},\ \bibinfo {pages} {3277}
  (\bibinfo {year} {1984})}\BibitemShut {NoStop}%
\bibitem [{\citenamefont {Yamamoto}\ and\ \citenamefont
  {Makimoto}(1974)}]{Yama74}%
  \BibitemOpen
  \bibfield  {author} {\bibinfo {author} {\bibfnamefont {S.}~\bibnamefont
  {Yamamoto}}\ and\ \bibinfo {author} {\bibfnamefont {T.}~\bibnamefont
  {Makimoto}},\ }\href@noop {} {\bibfield  {journal} {\bibinfo  {journal} {J.
  Appl. Phys.}\ }\textbf {\bibinfo {volume} {45}},\ \bibinfo {pages} {882}
  (\bibinfo {year} {1974})}\BibitemShut {NoStop}%
\bibitem [{\citenamefont {Popkov}\ \emph {et~al.}(1998)\citenamefont {Popkov},
  \citenamefont {Fehndrich}, \citenamefont {Lohmeyer},\ and\ \citenamefont
  {Dotsch}}]{Popk98}%
  \BibitemOpen
  \bibfield  {author} {\bibinfo {author} {\bibfnamefont {A.~F.}\ \bibnamefont
  {Popkov}}, \bibinfo {author} {\bibfnamefont {M.}~\bibnamefont {Fehndrich}},
  \bibinfo {author} {\bibfnamefont {M.}~\bibnamefont {Lohmeyer}}, \ and\
  \bibinfo {author} {\bibfnamefont {H.}~\bibnamefont {Dotsch}},\ }\href@noop {}
  {\bibfield  {journal} {\bibinfo  {journal} {Appl. Phys. Lett.}\ }\textbf
  {\bibinfo {volume} {72}},\ \bibinfo {pages} {2508} (\bibinfo {year}
  {1998})}\BibitemShut {NoStop}%
\bibitem [{\citenamefont {Udalov}\ \emph {et~al.}(2012)\citenamefont {Udalov},
  \citenamefont {Sapozhnikov}, \citenamefont {Karashtin}, \citenamefont
  {Gribkov}, \citenamefont {Gusev}, \citenamefont {Skorohodov}, \citenamefont
  {Rogov}, \citenamefont {Klimov},\ and\ \citenamefont {Fraerman}}]{Udal12}%
  \BibitemOpen
  \bibfield  {author} {\bibinfo {author} {\bibfnamefont {O.~G.}\ \bibnamefont
  {Udalov}}, \bibinfo {author} {\bibfnamefont {M.~V.}\ \bibnamefont
  {Sapozhnikov}}, \bibinfo {author} {\bibfnamefont {E.~A.}\ \bibnamefont
  {Karashtin}}, \bibinfo {author} {\bibfnamefont {B.~A.}\ \bibnamefont
  {Gribkov}}, \bibinfo {author} {\bibfnamefont {S.~A.}\ \bibnamefont {Gusev}},
  \bibinfo {author} {\bibfnamefont {E.~V.}\ \bibnamefont {Skorohodov}},
  \bibinfo {author} {\bibfnamefont {V.~V.}\ \bibnamefont {Rogov}}, \bibinfo
  {author} {\bibfnamefont {A.~Y.}\ \bibnamefont {Klimov}}, \ and\ \bibinfo
  {author} {\bibfnamefont {A.~A.}\ \bibnamefont {Fraerman}},\ }\href@noop {}
  {\bibfield  {journal} {\bibinfo  {journal} {Phys.\ Rev.\ B}\ }\textbf
  {\bibinfo {volume} {86}},\ \bibinfo {pages} {094416} (\bibinfo {year}
  {2012})}\BibitemShut {NoStop}%
\bibitem [{\citenamefont {Ford}\ and\ \citenamefont {Werner}(1978)}]{Ford78}%
  \BibitemOpen
  \bibfield  {author} {\bibinfo {author} {\bibfnamefont {G.~W.}\ \bibnamefont
  {Ford}}\ and\ \bibinfo {author} {\bibfnamefont {S.~A.}\ \bibnamefont
  {Werner}},\ }\href@noop {} {\bibfield  {journal} {\bibinfo  {journal} {Phys.\
  Rev.\ B}\ }\textbf {\bibinfo {volume} {18}},\ \bibinfo {pages} {6752}
  (\bibinfo {year} {1978})}\BibitemShut {NoStop}%
\bibitem [{\citenamefont {Lin}\ and\ \citenamefont {Chui}(2004)}]{Lin04}%
  \BibitemOpen
  \bibfield  {author} {\bibinfo {author} {\bibfnamefont {Z.}~\bibnamefont
  {Lin}}\ and\ \bibinfo {author} {\bibfnamefont {S.~T.}\ \bibnamefont {Chui}},\
  }\href@noop {} {\bibfield  {journal} {\bibinfo  {journal} {Phys.\ Rev.\ E}\
  }\textbf {\bibinfo {volume} {69}},\ \bibinfo {pages} {056614} (\bibinfo
  {year} {2004})}\BibitemShut {NoStop}%
\bibitem [{\citenamefont {Tarento}\ \emph {et~al.}(2004)\citenamefont
  {Tarento}, \citenamefont {Bennemann}, \citenamefont {Joyes},\ and\
  \citenamefont {de~Walle}}]{Tar04}%
  \BibitemOpen
  \bibfield  {author} {\bibinfo {author} {\bibfnamefont {R.-J.}\ \bibnamefont
  {Tarento}}, \bibinfo {author} {\bibfnamefont {K.-H.}\ \bibnamefont
  {Bennemann}}, \bibinfo {author} {\bibfnamefont {P.}~\bibnamefont {Joyes}}, \
  and\ \bibinfo {author} {\bibfnamefont {J.~V.}\ \bibnamefont {de~Walle}},\
  }\href@noop {} {\bibfield  {journal} {\bibinfo  {journal} {Phys.\ Rev.\ E}\
  }\textbf {\bibinfo {volume} {69}},\ \bibinfo {pages} {026606} (\bibinfo
  {year} {2004})}\BibitemShut {NoStop}%
\bibitem [{\citenamefont {Prosandeev}\ \emph {et~al.}(2008)\citenamefont
  {Prosandeev}, \citenamefont {Ponomareva}, \citenamefont {Korneev},\ and\
  \citenamefont {Bellaiche}}]{Pros08}%
  \BibitemOpen
  \bibfield  {author} {\bibinfo {author} {\bibfnamefont {S.}~\bibnamefont
  {Prosandeev}}, \bibinfo {author} {\bibfnamefont {I.}~\bibnamefont
  {Ponomareva}}, \bibinfo {author} {\bibfnamefont {I.}~\bibnamefont {Korneev}},
  \ and\ \bibinfo {author} {\bibfnamefont {L.}~\bibnamefont {Bellaiche}},\
  }\href@noop {} {\bibfield  {journal} {\bibinfo  {journal} {Phys.\ Rev.\
  Lett.}\ }\textbf {\bibinfo {volume} {100}},\ \bibinfo {pages} {047201}
  (\bibinfo {year} {2008})}\BibitemShut {NoStop}%
\bibitem [{\citenamefont {Landau}\ and\ \citenamefont
  {Lifshitz}(1984)}]{Landau8}%
  \BibitemOpen
  \bibfield  {author} {\bibinfo {author} {\bibfnamefont {L.~D.}\ \bibnamefont
  {Landau}}\ and\ \bibinfo {author} {\bibfnamefont {E.~M.}\ \bibnamefont
  {Lifshitz}},\ }\href@noop {} {\emph {\bibinfo {title} {Course of Theoretical
  Physics, Vol. 8: Electrodynamics of Continuous Media}}}\ (\bibinfo
  {publisher} {Butterworth - Heinemann, Oxford},\ \bibinfo {year}
  {1984})\BibitemShut {NoStop}%
\bibitem [{\citenamefont {Berger}\ and\ \citenamefont {Pufall}(1999)}]{Ber99}%
  \BibitemOpen
  \bibfield  {author} {\bibinfo {author} {\bibfnamefont {A.}~\bibnamefont
  {Berger}}\ and\ \bibinfo {author} {\bibfnamefont {M.~R.}\ \bibnamefont
  {Pufall}},\ }\href@noop {} {\bibfield  {journal} {\bibinfo  {journal} {J.
  Appl. Phys.}\ }\textbf {\bibinfo {volume} {85}},\ \bibinfo {pages} {4583}
  (\bibinfo {year} {1999})}\BibitemShut {NoStop}%
\bibitem [{\citenamefont {Dehesa-Martinez}\ \emph {et~al.}(2001)\citenamefont
  {Dehesa-Martinez}, \citenamefont {Blanco-Guttierez}, \citenamefont {Velez},
  \citenamefont {Diaz}, \citenamefont {Alvarez-Prado},\ and\ \citenamefont
  {Alameda}}]{Deh01}%
  \BibitemOpen
  \bibfield  {author} {\bibinfo {author} {\bibfnamefont {C.}~\bibnamefont
  {Dehesa-Martinez}}, \bibinfo {author} {\bibfnamefont {L.}~\bibnamefont
  {Blanco-Guttierez}}, \bibinfo {author} {\bibfnamefont {M.}~\bibnamefont
  {Velez}}, \bibinfo {author} {\bibfnamefont {J.}~\bibnamefont {Diaz}},
  \bibinfo {author} {\bibfnamefont {L.~M.}\ \bibnamefont {Alvarez-Prado}}, \
  and\ \bibinfo {author} {\bibfnamefont {J.~M.}\ \bibnamefont {Alameda}},\
  }\href@noop {} {\bibfield  {journal} {\bibinfo  {journal} {Phys.\ Rev.\ B}\
  }\textbf {\bibinfo {volume} {64}},\ \bibinfo {pages} {024417} (\bibinfo
  {year} {2001})}\BibitemShut {NoStop}%
\bibitem [{\citenamefont {Zharov}\ and\ \citenamefont {Kurin}(2007)}]{Kur07}%
  \BibitemOpen
  \bibfield  {author} {\bibinfo {author} {\bibfnamefont {A.~A.}\ \bibnamefont
  {Zharov}}\ and\ \bibinfo {author} {\bibfnamefont {V.~V.}\ \bibnamefont
  {Kurin}},\ }\href@noop {} {\bibfield  {journal} {\bibinfo  {journal} {J.
  Appl. Phys.}\ }\textbf {\bibinfo {volume} {102}},\ \bibinfo {pages} {123514}
  (\bibinfo {year} {2007})}\BibitemShut {NoStop}%
\bibitem [{\citenamefont {Landau}\ and\ \citenamefont
  {Lifshitz}(1975)}]{Landau2}%
  \BibitemOpen
  \bibfield  {author} {\bibinfo {author} {\bibfnamefont {L.~D.}\ \bibnamefont
  {Landau}}\ and\ \bibinfo {author} {\bibfnamefont {E.~M.}\ \bibnamefont
  {Lifshitz}},\ }\href@noop {} {\emph {\bibinfo {title} {Course of Theoretical
  Physics, Vol. 2: The Classical Theory of Fields}}}\ (\bibinfo  {publisher}
  {Butterworth - Heinemann, Oxford},\ \bibinfo {year} {1975})\BibitemShut
  {NoStop}%
\end{thebibliography}%

\end{document}